\RequirePackage{luatex85} 
\documentclass[preprint,12pt,3p]{elsarticle}

\usepackage{latexsym}
\usepackage{dcolumn}
\usepackage{supertabular,multirow}
\usepackage{epsfig}
\usepackage{bm}
\usepackage{subfigure}
\usepackage{color}
\usepackage{hyperref}
\usepackage{tikz,siunitx}
\usepackage{amssymb,amsfonts,amsmath}
\usepackage{euscript}
\usepackage{graphics}
\usepackage{setspace}
\usepackage{graphicx}
\usepackage{hyperref}
\usepackage{xcolor}
\usepackage{siunitx}
\usepackage[normalem]{ulem}
\usepackage[colorinlistoftodos]{todonotes}
\usepackage{url}

\definecolor{brown}{rgb}{0.59, 0.29, 0.0}
\definecolor{orange}{RGB}{255,127,0}
\definecolor{brightube}{rgb}{0.48, 0.12, 0.21}

\graphicspath{{Figs/}}

\newcommand{\deleted}[1]{}

\newcommand{\eqn}[1]{\begin{align}#1\end{align}}
\newcommand{\bs}[1]{\boldsymbol{#1}}

\newcommand{\pare}[1]{\left( #1 \right) }
\newcommand{\corchete}[1]{\left[ #1 \right]}
\newcommand{\fr}[2]{\frac{#1}{#2}}
\newcommand{\wtil}[1]{\widetilde{#1}}
\newcommand{\mc}[1]{\mathcal{#1}}

\newcommand{\tex}[1]{\mbox{\scriptsize{#1}}}

\def\dt{\Delta t}
\def\dd{\mathrm{d}}

\def\bbf{\bs{f}}
\def\bF{\bs{F}}
\def\bG{\bs{G}}
\def\bI{\bs{I}}

\def\bK{\bs{K}}

\def\bM{\bs{M}}

\def\bq{\bs{q}}

\def\br{\bs{r}}

\def\bu{\bs{u}}
\def\bU{\bs{U}}
\def\bv{\bs{v}}

\def\bx{\bs{x}}

\def\bzero{\bs{0}}

\def\blambda{\bs{\lambda}}

\def\bmK{\bs{\mc{K}}}

\journal{}

\begin{document}

\begin{frontmatter}

\title{Mapping flagellated swimmers to surface-slip driven swimmers}

\author[inst1]{Harinadha Gidituri}
\author[inst2]{G\"{o}kberk Kabacao\u{g}lu}
\author[inst1,inst3,inst4]{Marco Ellero}
\author[inst1]{Florencio Balboa Usabiaga}

\affiliation[inst1]{organization={BCAM - Basque Center for Applied Mathematics},
  city={Mazarredo 14, Bilbao},
  postcode={E48009}, 
  country={Basque Country - Spain}}

\affiliation[inst2]{organization={Department of Mechanical Engineering, Bilkent University},
  city={Ankara},
  postcode={06800},  
  country={Turkey}}

\affiliation[inst3]{organization={Ikerbasque, Basque Foundation for Science},
  city={Maria Diaz de Haro 3, Bilbao},
  postcode={E48013},  
  country={Basque Country - Spain}}

\affiliation[inst4]{organization={Zienkiewicz Center for Computational Engineering (ZCCE), Swansea University},
  city={Bay Campus, Swansea SA1 8EN},
  country={UK}}

\begin{abstract}
Flagellated microswimmers are ubiquitous in natural habitats. Understanding the
hydrodynamic behavior of these cells is of paramount interest, owing to their applications in
bio-medical engineering and disease spreading. Since the last two decades, computational efforts
have been continuously improved to accurately capture the complex hydrodynamic behavior of these model systems.
However, modeling the dynamics of such swimmers with fine details is computationally expensive 
due to the large number of unknowns and the small time-steps required to solve the equations.
In this work we propose a method to map fully resolved flagellated microswimmers to coarse,
active slip driven swimmers which can be simulated at a reduced computational cost.
Using the new method, the slip driven swimmers move with the same velocity, to machine precision, as the flagellated swimmers and generate
a similar flow field with a controlled accuracy.
The method is validated for swimming patterns near a no-slip boundary, interactions between swimmers and scattering with large obstacles.
\end{abstract}

\begin{keyword}
Microswimmers \sep Squirmer \sep Stokes flow \sep Fluid-structure interactions
\end{keyword}

\end{frontmatter}

\tableofcontents

\section{Introduction}
\label{sec:introduction}

An assortment of microbes have evolved to use cilia and flagella for swimming.
For example, \emph{Paramecia} are covered by a carpet of cilia that beat with an asymmetric stroke that propel the organisms forward \cite{Narematsu2015};
many bacteria rotate a small number of flagella to use them as a screw and advance in the fluid \cite{Drescher2011,Kuehn2017};
Protists show an incredible diversity in the morphology and arrangement of flagella that they use to swim and generate feeding currents \cite{Nielsen2021}.
The taxis of all these organisms can be studied with computational fluid dynamics (CFD) methods.
However, there is always a trade-off between accuracy and computational cost which makes it difficult to determine the best numerical method to study a given problem.

Well-resolved models to simulate microswimmers require solving the hydrodynamic interactions between the swimmers body and the flagella or cilia
while enforcing the attachment constraints to the body.
As the Reynolds number at the relevant scales is small the schemes can use the Stokes equations to describe the hydrodynamic interactions.
The Stokes equations are an elliptic PDE, therefore, the flow is uniquely determined by the boundary conditions \cite{Kim1991, Pozrikidis1992}.
Such property is a numerical advantage as it allows to use approaches like the boundary integral method that only discretizes the swimmer
surface but not the fluid domain,
since the flow can be computed by the action of the Green's functions of the Stokes equation \cite{Corona2017, Corona2018}.
However, when the discretization includes the flagella and cilia the number of unknowns in the equations grow fast \cite{Higdon1979, Das2018, Walker2019, Westwood2021, Usabiaga2022}.
Therefore, performing many body simulations can be expensive.
An additional challenge for well-resolved models is the time step size limitation.
The time step has to be, at least,  smaller than the characteristic beating period of the cilia and flagella.
For many organisms that time scale is on the order of $0.01\,\si{s}$ while their swimming speeds can be around $10\,\si{\mu m/s}$.
Therefore, at least a $100$ steps are required for a swimmer to move one body length ($\sim 10\,\si{\mu m}$) \cite{VelhoRodrigues2021}.
When the flexibility of the appendages is modeled additional care and time steps restrictions are necessary \cite{Tornberg2004, Maxian2021, Maxian2022, Schoeller2021}.

To reduce the computational cost more simplified models have been used.
The most commons are the ones based on multipole expansions and those including an active slip on the swimmers surface.
Both approaches require less unknowns than well-resolved methods and they can use much larger time steps.
The models based on multipole expansions describe the flow field generated by the microswimmers as a sum of Stokes multipoles.
For force-free microswimmers the lowest relevant multipole is the force dipole  \cite{Hernandez-Ortiz2005},
which is enough to distinguish between pushers as the bacteria \emph{e. coli} \cite{Drescher2011}
and pullers as some choanoflagellates \cite{Smith2009} or algae \cite{Drescher2010}.
Most minimal models do not include higher order multipoles but some of them include torque-dipoles or stresslets \cite{Delmotte2015c,Ishimoto2020a}.
These models have been very successful to explain the instability of suspensions of pushers \cite{Saintillan2008, Saintillan2008a, Lushi2018}
and even the scattering of swimmers with obstacles \cite{Spagnolie2015a, Contino2015, Lushi2017}.
However, they present important limitations.

First, it is generally difficult to improve them by including higher order multipoles;
and perhaps, a bigger problem is that they generate flows that, do not obey the boundary condition on the swimmers surfaces.
This limitation can be ameliorated by including non-pairwise terms, such as stresslets, but that reduces the simplicity of the method
and again it is hard to include high order multipoles \cite{Swan2011, Delmotte2015c, Elfring2022}.
These limitations restrict the use of these methods to study suspensions in the dilute regime
or the scattering with obstacles with only simple geometries for which the image system of the Stokes equations is known.

The other common approach is to model microswimmers as rigid particles with an active slip on their surfaces.
The effective slip models the thrust that is generated by a carpet of underresolved cilia, as introduced in the squirmer model by Blake in the 1970s \cite{Blake1971a}.
The slip can be time-dependent \cite{Guo2021} although most models use constant slips.
This approach has been used in multitude of works to study different phenomena like
interactions between swimmers \cite{Llopis2010}, sedimentation \cite{Scagliarini2022}, swimming near walls \cite{Spagnolie2012, Ishimoto2013} or interfaces \cite{Gidituri2022}.
However, the optimum choice of the slip to match a specific swimmer has not been explored in detail.
Most works consider axisymmetric slips expanded as a Legendre polynomial truncated at the second order;
but a generalization where a rotlet term is included was presented in Ref. \cite{Fadda2020},
very nice work exploring the optimal time-dependent slip for propulsion was presented in Ref. \cite{Guo2021} and
a comparison between ciliated and squirmer swimmers was provided in Ref. \cite{Ito2019}.

In the present work we propose a novel scheme,
inspired by the method of fundamental solutions \cite{Bogomolny1985, Liu2016a, Stein2022},
to generate an active slip to match the flow field generated by any force-free swimmer.
Our scheme guarantees that the mapped swimmer generates a similar flow field, with a controllable accuracy, as the original swimmer
and that it swims with the same velocity, to machine precision, in bulk.
At the same time the computational is cost is much lower than in fully resolved methods as it is not necessary to resolve
the time scale associated with the stroke of cilia and flagella.
Thus, this numerical method could be useful in parametric, optimization or uncertainty quantification problems.
Any method can be used to simulate the original swimmer.
Our mapping scheme only requires the swimming velocity and the flow generated on a shell bounding the swimmer.
To simulate the active slip swimmer any boundary-integral-like method can be used and we choose the rigid multiblob method as described in Sec.\ \ref{sec:rmb}.
In section \ref{sec:map} we describe our mapping scheme and present numerical results in Sec. \ref{sec:results}.
Finally, the conclusions are presented in Sec. \ref{sec:conclusions}.

\section{Hydrodynamic equations}
\label{sec:rmb}
In this section we present our model for simulating the motion of rigid bodies in inertialess viscous flows governed by the Stokes equations
\begin{align}
\label{eq:Stokeseqn}
- \nabla \pi + \eta \nabla^2 {\bs{v}} = \bzero, \\
\nabla \cdot {\bs{v}} = 0,
\end{align}
where $\pi$ and $\bv$ are the fluid pressure and velocity and $\eta$ its viscosity.
The configuration of a rigid body $p$, $\mathcal{B}_p$, is represented using two parts: 1) the location of the tracking point ${\bs {q}_{p}}$ and 2) the orientation of the body ${\bs{\theta}_{p}}$.
The linear and angular velocities are denoted by ${\bs{u}_{p}}$ and ${\bs{\omega}_{p}}$.
Externally applied forces and torques are represented by ${\bs{f}_{p}}$ and ${\bs{\tau}_{p}}$. In a compact notation we can write as ${\bs{U}_{p}} = ({\bs{u}_{p}}, {\bs{\omega}_{p}})$ and ${\bs{F}_{p}} = ({\bs{f}_{p}}, {\bs{\tau}_{p}})$.
The fluid obeys the no-slip boundary condition at the bodies surface.
In the case of particles with an active slip, (eg. phoretic particles or squirmers)
an additional slip is accounted for.
We write the slip boundary condition on one body as,
\begin{equation}
  \label{eq:no-slip}
  \bv(\br) = {\bs{u}_{p}} + {\bs{\omega}_{p}}\times ({\bs{r}}-{\bs {q}_{p}} ) + \bu_s(\br)  \;\text{ for } \br \in \partial \mathcal{B}_p
\end{equation}
where $\partial \mathcal{B}_p$ is the surface of body $p$ and $\bu_{s}$ is the active slip. 
We can write the above equation in short form as
\begin{equation}
  \label{eq:no-slip_short} 
  \bv(\br) = \bmK_p \bU_p + \bu_s(\br)  \;\text{ for } \br \in \partial \mathcal{B}_p,
\end{equation}
where the linear operator $\bmK_p$ transforms rigid body velocity into surface velocities.
For body $p$ the balance between the hydrodynamic traction, $\blambda$, and the external forces and torques is given by
\begin{align}
  \label{eq:force-balance}
  \int_{\partial B_p} \blambda(\br)\; \dd S_{\br} \ = \bbf_p, \\
  \label{eq:torque-balance}
  \int_{\partial B_p} (\br  -{\bs {q}}_p) \times \blambda(\br) \; \dd S_{\br} \  = {\bs{\tau}}_p.
\end{align}

\subsection{Rigid multiblob method} 
We use the rigid multiblob method to discretize and solve the hydrodynamic equations \cite{Usabiaga2016}.
In this method the rigid bodies surfaces are discretized with a finite number of blobs, $N_b$, with position $\br_i$ as in Fig.\ \ref{fig:sketch}.
Once the body $p$ is discretized, the balance of force and torque  becomes
\begin{align}
  \label{eq:force-balance-discrete}
  \sum_{i\in \mathcal{B}_{p}} \blambda_{i} = \bbf_p, \\
  \label{eq:torque-balance-discrete}
  \sum_{i\in \mathcal{B}_{p}} (\bs{r}_i  -{\bs {q}}_p) \times {\bs{\lambda}}_i  = {\bs{\tau}}_p,
\end{align}
where $\blambda_i$ is the hydrodynamic traction acting on the blob $i$.
The slip condition is evaluated at each blob $i$,
\begin{equation}
  \label{eq:no-slip-discrete}
  \bv(\br_i) = \sum_{j} \bM_{ij} \blambda_j =  {\bs{u}_{p}} + {\bs{\omega}_{p}}\times ({\bs{r}_{i}}-{\bs {q}_{p}} ) + \bu_{s,i} \; \text{ for } i\in \mathcal{B}_{p}.
\end{equation}
The mobility matrix $\bM_{ij}$ gives the hydrodynamic interaction between any two blobs, $i$ and $j$, of radius $a_i$ and $a_j$.
The rigid multiblob method uses as mobility the Rotne-Prager-Yamakawa (RPY) tensor, a regularization of the Stokes kernel.
It can be written as \cite{Wajnryb2013}
\begin{equation}
  \label{eq:rpy}
  \bM_{ij} = \fr{1}{(4\pi a_i a_j)^2} \int \delta (|{\bs{r'-r}_{i}}|- a_i) {\bs{G({\bs{r',r''}}})}\delta (|{\bs{r''-r}_{j}}|- a_j) \, \dd^3r' \dd^3r'' ,
\end{equation}
where $\bG(\br,\br')$ is the Green's function of the Stokes equation and $\delta(\br)$ the Dirac delta function.
The RPY mobility is always positive definite which makes the rigid multiblob method very robust and easy to use.
There are fast methods to compute the action of the RPY mobility in bulk \cite{Liang2013}, near an infinite flat wall \cite{Yan2018a}
and in periodic domains \cite{Fiore2017}.

The equations \eqref{eq:force-balance-discrete}-\eqref{eq:no-slip-discrete} form a linear system.
We can write it in a compact form by first defining the $3N_b \times 6$ matrix $\bK_p$,
a discretization of the linear operator $\bmK_p$, with block elements
\eqn{
  \bK_{p,i} = \begin{cases}
    \corchete{\bI_{3\times 3} \;\; -(\br_i - \bq_p)^{\times}}
  \end{cases}
}
where $\bI_{3\times 3}$ is the $3\times 3$ identity matrix and $(\br_i - \bq_p)^{\times} \bx = (\br_i - \bq_p) {\times} \bx$ for any vector $\bx$.
Then, the system of equations for a suspension of $N$ bodies can be written as
\eqn{ 
  \label{eq:linear_system}
  \begin{bmatrix}
    \bM & -\bK \\
    -\bK^T & \bzero \\
  \end{bmatrix}
  \begin{bmatrix}
    \blambda \\ \bU
  \end{bmatrix}
  =
  \begin{bmatrix}
    \bu_s \\ -\bF
  \end{bmatrix},
}
where unscripted vectors refer to the composite vector formed by the variables of all the bodies, e.g.\ $\bU = \{\bU_p\}_{p=1}^N$,
and the matrix $\bK$ is block-diagonal, i.e. $\bK=\text{Diag}\{\bK_p\}_{p=1}^N$.
Note that all bodies in a suspension are hydrodynamically coupled through the dense matrix $\bM$.
The linear system \eqref{eq:linear_system} can be solved with a Krylov method such as GMRES \cite{Usabiaga2016}.
Once the velocity $\bU$ has been found the equations of motion can be integrated in time \cite{Sprinkle2017}.
Also, once the force on the blobs, $\blambda$, has been found it is possible to compute the flow anywhere in the fluid as
\eqn{
  \label{eq:flow}
  \bv(\br_i) = \sum_j \bM_{i j} \blambda_j,
}
where the sum runs over all the blobs in the system.
The test blob at $\br_i$ is taken to have radius $a_i=0$ in \eqref{eq:rpy} to measure the pointwise  velocity of the flow.
The rigid multiblob method has been extended to simulate flagellated swimmers and we will use it in Sec.\ \ref{sec:results}.
A detailed description can be found elsewhere \cite{Usabiaga2022}.

\section{Swimmer map algorithm}
\label{sec:map}

\begin{figure}
  \begin{center}
    \includegraphics[width = 0.89 \columnwidth]{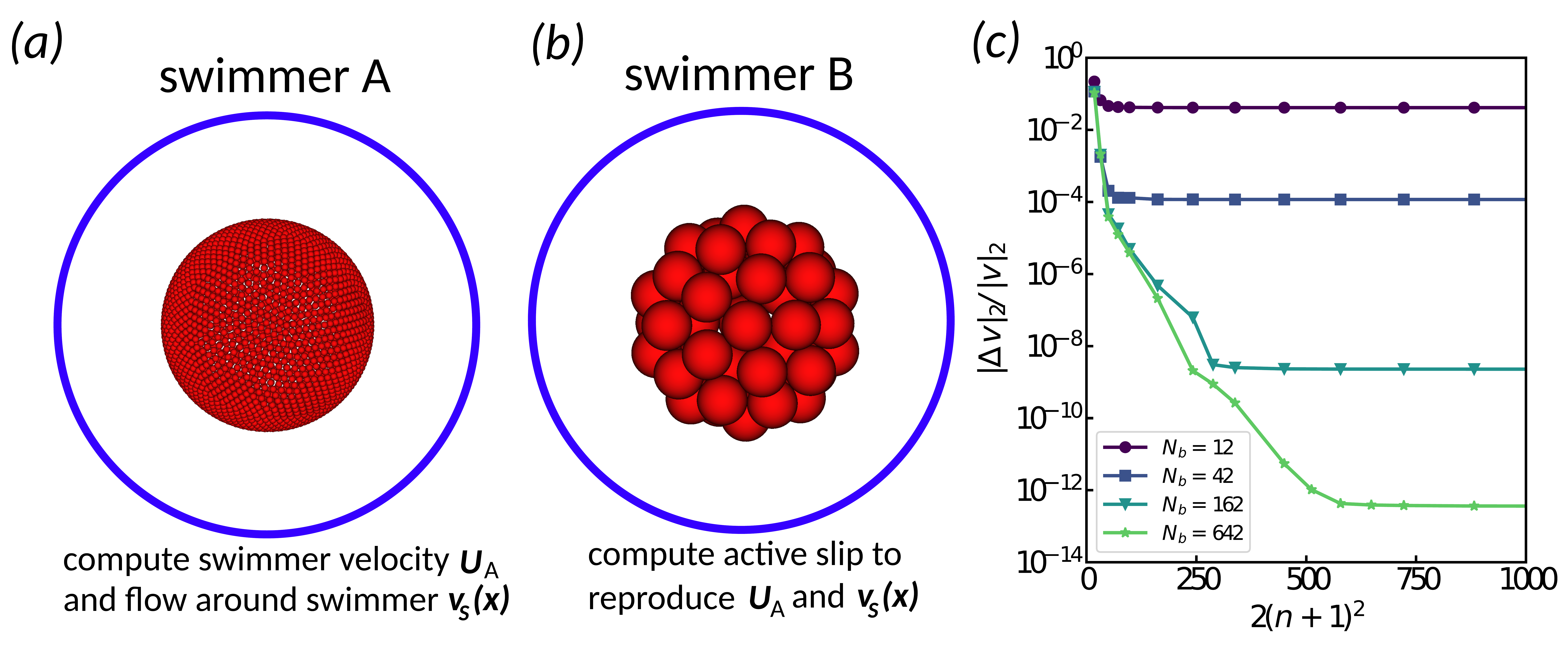}
    \caption{{\bf (a,b)} Sketch of the mapping scheme: the flow generated by the swimmer A is computed on the blue shell.
      Then an optimum slip is computed for swimmer B, discretized with blobs, to generate the same flow and swimming speed as for swimmer A.
      {\bf (c)} Flow convergence for the squirmer case.
      We map a squirmer discretized with $N_b=2562$ blobs with lower resolution discretizations with $N_b=12,\,42,\,162\text{ and }642$ blobs.
      We compute the relative velocity error on a well-resolved shell ($n=32$) for slips calculated with the flow on shells with different number of points, $2(n+1)^2$.
      The error decreases with $n$ while the linear system \eqref{eq:least_squares} is underdetermined, around that point the error reaches a plateau.
      The plateau error can be decreased by using a higher number of blobs, $N_b$, to discretize the squirmer.
    }
    \label{fig:sketch}
  \end{center}
\end{figure}

We describe here our algorithm to determine the optimum slip to map flagellated swimmers to coarse surface slip driven swimmers.
Our algorithm guarantees that the swimming velocity of the flagellated swimmer will be matched to machine precision while the flow field will be matched
with a controllable accuracy.
For simplicity, in this section we label the first and second swimmers as A and B respectively.
Swimmer A should be force-torque free during calculations.
After the slip is determined the swimmers can be subject to external forces such as gravity or steric interactions with obstacles.
Unless otherwise stated, swimmer A will be a flagellated organism like a bacterium or a paramecium but the same scheme can be used for
other kinds of swimmers such as self-propelled phoretic particles.

The inputs to the scheme are the swimming velocity of swimmer A, $\bU_A$, the flow that generates on a shell centered on the swimmer, $\bv_S$,
and a discretization of swimmer B see Fig.\ \ref{fig:sketch}b.
Any numerical method that solves the Stokes equation can be used for this step.
The flow on the shell should be well resolved, thus, we compute the flow on a Chebyshev-Fourier grid of order $n$ with $2(n+1)^2$ points \cite{Veerapaneni2011}.
If we are interested in determining a time dependent slip, the swimming velocity and the surrounding flow are instantaneous.
If we aim at the average dynamics, they are time-averaged values in the body frame of reference.
The only difference for the scheme is that with time-dependent slips the optimization problem must be solved several times.
Average slips allow much larger step sizes in the simulation of active slip driven swimmers and therefore we will consider them in the examples in Sec.\ \ref{sec:results}.
Some swimmers, like the algae \emph{Chlamydomonas reinhardtii}, generate time-dependent flows \cite{Guasto2010}
and the time variation can be crucial to reproduce the dynamics of some systems \cite{Delmotte2015c, Walker2022}.
Our algorithm can tackle both situations.

Given the inputs we compute the surface density force on the swimmer B that generates the same flow on the shell, $\bv_S$, as the swimmer A.
In the Stokes equations the flow is completely determined by the boundary conditions,
therefore, if the flow on the shell is matched, the exterior flow will be the same for both swimmers \cite{Kim1991,Pozrikidis1992}.
For the rigid multiblob method the inverse problem to find the blobs' forces reduces to solving the following linear problem
\eqn{
  \label{eq:least_squares}
  \left[\begin{array}{c}
      \bM_{SB} \\
      w_1 \bK_B^T
    \end{array}\right] \blambda_B =
  \left[\begin{array}{c}
      \bv_S \\
      \bzero
    \end{array}\right].
}
In the first equation of the linear system \eqref{eq:least_squares} the mobility $\bM_{SB}$ couples the force acting on the blobs to the flow on the shell
as defined by \eqref{eq:flow}. 
The second equation is a constraint to guarantee that the swimmer B is indeed force and torque free.
This constraint is not necessary if the flow on the shell is well resolved and the swimmer B is well discretized.
However, we have found that when the swimmer B is only discretized with a small number of blobs the constraint is necessary.
The weight $w_1$ controls the importance given to the constraint; we use
\eqn{
  \label{eq:weight}
  w_1 = \fr{1}{\eta a|\wtil{\blambda}_B|}\fr{|\bK_B^T \wtil{\blambda}_B|}{\epsilon},
}
where the blob forces $\wtil{\blambda}_B$ are the solution to the unconstrained resistance problem $\bM_{SB} \wtil{\blambda}_B = \bv_{S}$,
the prefactor $1/(\eta a|\wtil{\blambda}_B|)$ sets the units and $\epsilon$ is a tight tolerance.
We typically use $\epsilon \in \left[10^{-14}, 10^{-6}\right]$.
Therefore, the constraint in \eqref{eq:least_squares} is only important if the solution of the unconstrained resistance problem was not force-torque free
to  the desired accuracy.
The linear system \eqref{eq:least_squares} and the unconstrained resistance problem are not necessarily square,
as in general, the number of blobs in swimmer B and the number of points on the shell can be different.
Therefore, we solve these systems in the least square sense using a singular value decomposition (SVD).
The SVD is also backward stable solver which allow us to find an accurate solution to \eqref{eq:least_squares} even
when the condition number of the linear system is very large ($\sim 10^{17}$) \cite{Yan2018, Stein2022}.

Once the force on the blobs is determined we compute the active slip directly from the slip condition \eqref{eq:no-slip} 
\eqn{
  \label{eq:active-slip}
  \bu_s = -\bK_{B} \bU_A + \bM \blambda_B.
}
The operators $\bK_B$ and $\bM$ are defined for the swimmer B but we use the known swimming velocity of swimmer A, $\bU_A$.
This trivial equation guarantees that the swimmer B will have the same velocity, to machine precision, as swimmer A.

\subsection{Validation}
\label{sec:validation}
To validate our scheme we show here a simple example and study more physical interesting problems in Sec.\ \ref{sec:results}.
In this test we show how a low resolution model can recover the flow generated by a swimmer and how the accuracy depends on the resolution.
We use as swimmer A a spherical squirmer of radius $R=1$ with tangential slip \cite{Blake1971a, Scagliarini2022}
\eqn{
  u_{s,\theta} = B_1 \sin \theta + \fr{B_2}{2} \sin 2\theta,
}
where $\theta$ is the polar angle defined on the surface and we set $B_1=B_2=1$.
We discretize the squirmer with $N_b=2562$ blobs and compute its velocity and the flow on a grid of order $n$ and $2(n+1)^2$ points
defined on a shell of radius $R_{\text{shell}}=4$.
As coarser swimmer B we use lower resolution discretizations with $N_b=12,\,42,\,162\text{ and } 642$ blobs.
For the four cases we compute the optimum slip using grids of different order $n$ to discretize the shell.
By construction the mapped swimmer recover the swimming velocity of swimmer A to machine precision.
To quantify the accuracy of the flow, once the optimum slip is computed, we compute the flow generated by the swimmer B on a shell discretized with a high-order grid
($n=32$ and $2(n+1)^2=2178$) and compare with the flow generated by swimmer A.
We show the relative errors in Fig.\ \ref{fig:sketch}c.
We can see that while the linear system \eqref{eq:least_squares} is underdetermined the error decreases with $n$.
For overdetermined systems the error reaches a plateau whose magnitude depends on the resolution used to discretized swimmer B.

The rigid multiblob has a slow convergence with the number of blobs, thus, the near flow will always have a large error,
as we will show in Sec.\ \ref{sec:results}, unless a very large number of blobs is used.
However, although we have delineated our algorithm for the rigid multiblob method it can be trivially adapted to use any boundary
integral method that allows for surface slips \cite{Bao2018, Corona2018}.
With a spectrally accurate boundary method it should be possible to match also the near field to an arbitrary precision with a reasonable number of points.
In fact, Stein et al. used similar ideas to develop a near field quadrature scheme for boundary integral methods that could be adapted to solve for the optimum slip
\cite{Stein2022, Young2021}.

\subsection{Computational cost}
\label{sec:cost}
The computational cost of simulating a suspension of swimmers depends on the time step size used to
integrate the equations of motion and the cost to solve a Stokes problem each time step.
The computational cost to solve one Stokes problem is linear in the number of degrees of freedom
when one use fast methods to compute the hydrodynamic interactions \cite{Yan2018, Yan2018a} and a preconditioned iterative solver \cite{Usabiaga2016, Usabiaga2022}.
Thus, the overall $\text{Cost} \sim (T / \dt) N_b$ where $T$ and $\dt$ are the total simulation time and the discrete time step size and $N_b$ is the number of degrees of freedom in the system. 

The mapping scheme can help to increase $\dt$ and reduce $N_b$.
The cost reduction is problem-dependent but we give here some characteristic numbers.
The swimming speed of \emph{E. coli} is of the order of $20\,\si{\mu m / s}$ while their flagella rotate with a period of $0.01\,\si{s}$ \cite{Darnton2007}.
When one models the bacteria as a flagellated swimmer it is necessary to use about 10 time steps per flagella rotation to solve the dynamics accurately,
thus, the bacteria would advance $0.02\,\si{\mu m}$, or about $1\%$ or their body length, per time step.
The mapping scheme presented here allows to increase the time step at least by a factor 10.
For ciliated swimmers the speed-up can be larger.
A swimmer covered by cilia can move up to one cilia length per beating and it is necessary about 100 time step sizes to resolved the cilia beating \cite{Omori2020, Westwood2021}.
Thus, if the cilia length is just a fraction of the swimmer size, e.g.\ one tenth, it is necessary about 1000 time steps size to advance the swimmer one body length.
The mapping scheme presented here allows to increase the time step at least by a factor 100.

The change in the number of degrees of freedom is also problem-dependent.
The results in Sec.\ \ref{sec:validation} show that for a simple shaped swimmer reducing the degrees of freedom by about a factor 10
allows to recover the flow two diameters away from the swimmer with a relative error of $\sim 10^{-8}$, see also Sec.\ \ref{sec:two_squirmers}.
To model complex shaped swimmers one may need to increase the number of degrees of freedom to generate an envelope that accurately covers the swimmer,
see for example the bacterium in \ref{sec:bacteria_flow}.
In that case the computational savings would come only from the larger time steps sizes.
However, it is still possible to use a coarser mesh by accepting a lower accuracy near the swimmer.

\section{Numerical results}
\label{sec:results}
To validate our numerical method we conduct several tests on microswimmer locomotion and flows in bulk and near boundaries.
First, we compare the flow generated by two nearby squirmers using a multipole expansion approximation and our method
in Sec.\ \ref{sec:two_squirmers}.
Then, we validate the flow field generated by a slip driven swimmer that maps a flagellated bacterium in Sec.\ \ref{sec:bacteria_flow}.
We also compare the trajectories of flagellated swimmers near obstacles;
near an infinite wall in Sec.\ \ref{sec:bacteria_wall} and near a spherical obstacle in Sec.\ \ref{sec:bacteria_scattering}.
Finally, we compare the trajectory of two flagellated swimmers with the trajectory of two slip driven swimmers in Sec.\ \ref{sec:two_bacteria}.
In all cases we obtain a good agreement between the dynamics of the mapped and original microswimmers.

\subsection{Nearby squirmers}
\label{sec:two_squirmers}

\begin{figure}
  \begin{center}
    \includegraphics[width=0.8 \columnwidth]{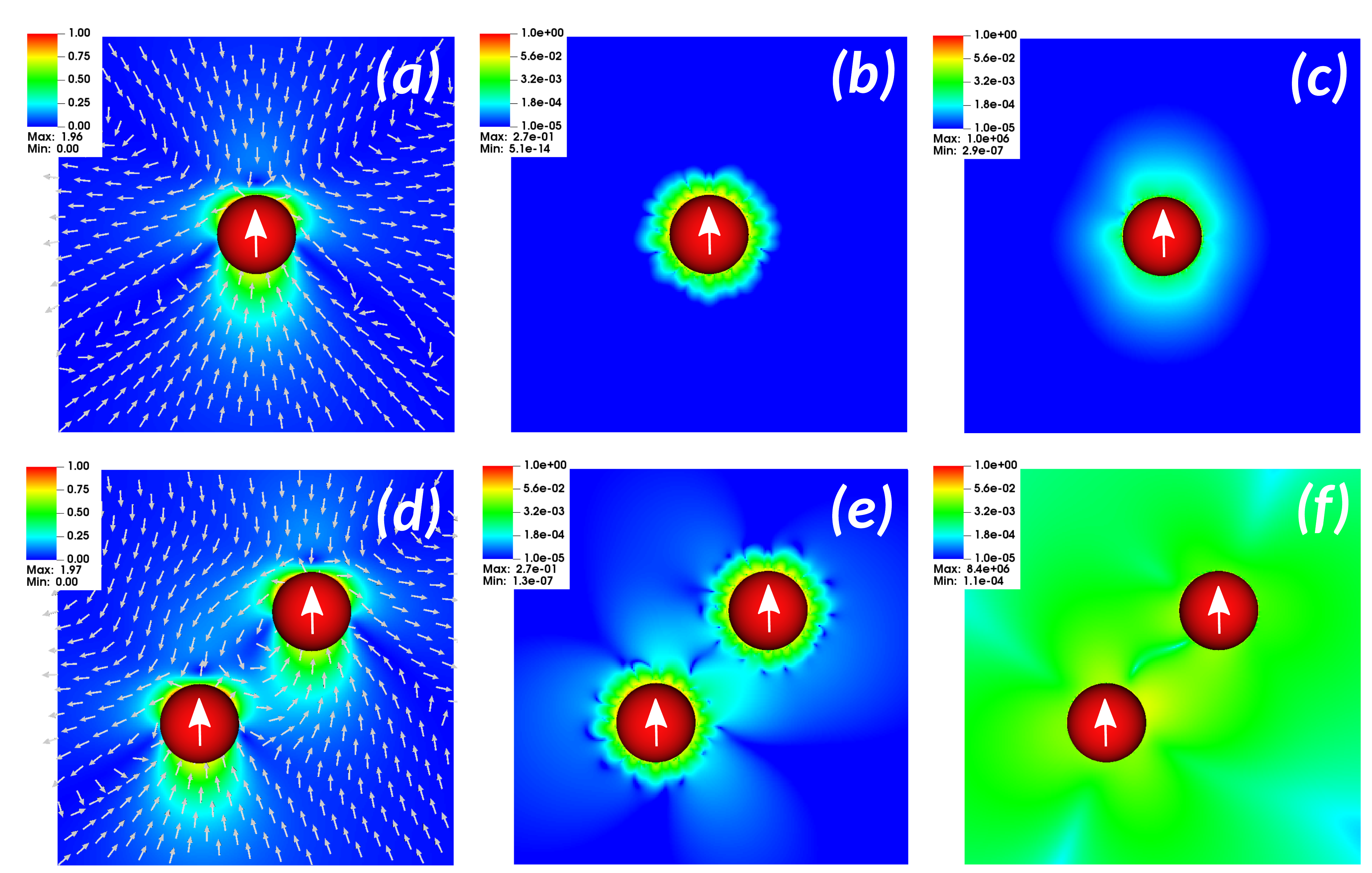}    
    \caption{Flow around one or two nearby squirmers (a,d) simulated with a discretization with $N_b=2562$ blobs.
      The maximum flow speed is $v_{\text{flow}} \approx 1$ and decays like $\sim 1/r^2$ with the distance to the swimmers.
      The panels (b,e) show the flow difference between the high resolution simulation and other with  a lower resolution of $N_b=162$ blobs.
      The panels (c,f) show the flow difference between the high resolution simulation and a simulation using a multipole expansion up to the octupole order.     
    }
    \label{fig:two_squirmers}
  \end{center}
\end{figure}

\begin{figure}
  \begin{center}
    \includegraphics[width=0.5 \columnwidth]{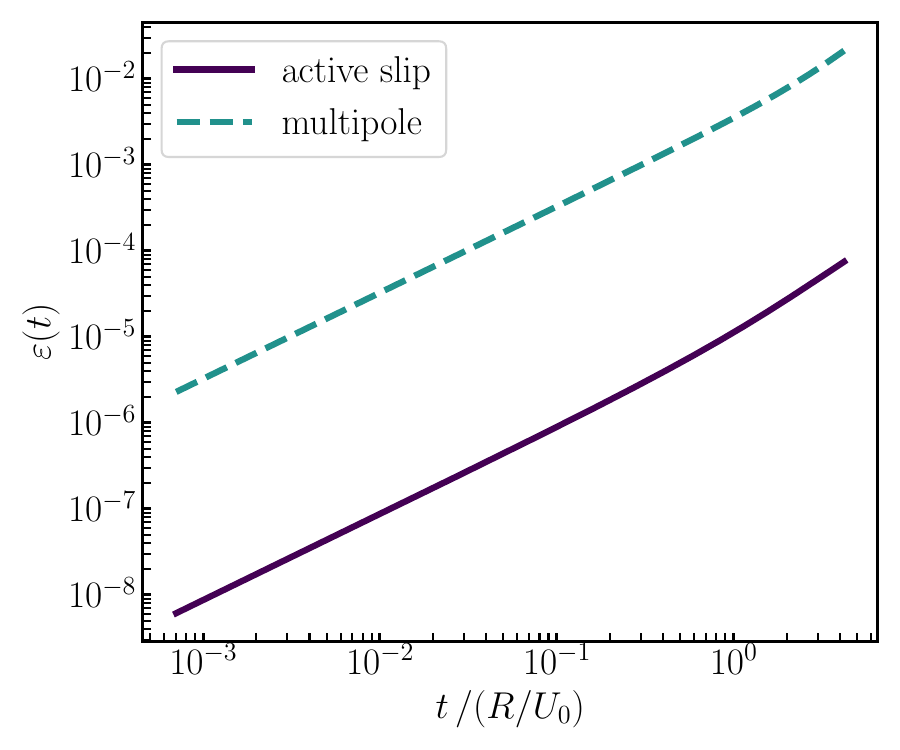}
    \caption{
      Trajectory difference between the well-resolved squirmers of Fig.\ \ref{fig:two_squirmers}d and those modeled with $N_b=162$ blobs or with a multipole method.
      The difference is defined as $\varepsilon(t) = |\bq_{\tex{modeled}}(t) - \bq_{\tex{well-resolved}}(t)|$ for one of the squirmers in the pair,
      where $\bq_{\tex{modeled}}(t)$ refers to the position of the squirmer modeled with an active slip and $N_b=162$ blobs (continuous line) or a multipole method (dashed line).
    }
    \label{fig:two_squirmers_trajectory}
  \end{center}
\end{figure}

Here we compare the flow generated by one or two squirmers modeled as slip driven swimmers or as a multipole expansion of fundamental solutions \cite{Spagnolie2012}.
This example shows that the multipole method fails to model the suspension of swimmers beyond the dilute regime accurately;
whereas our method is accurate at the flow around the swimmers even when hydrodynamic reflections are important.
First, we compute the flow around a single squirmer as the one in the previous section ($B_1=B_2=1$ and $N_b=2562$).
Then, we compare the difference in the flow generated by a squirmer modeled as a slip driven swimmer with $N_b=162$ blobs or with a multipole method
up to the octupole order \cite{Spagnolie2012}.
The top panels of Fig.\ \ref{fig:two_squirmers} show that in both cases the error is below $10^{-3}$ one radius away from the swimmer.
Thus, the multipole method is as accurate, at a lower computational cost, than the slip driven swimmer model to simulate single swimmers in bulk.
However, the results are different when considering multiple swimmers.
In the bottom panels of Fig.\ \ref{fig:two_squirmers}
we compare the flow difference for two nearby squirmers modeled as the same slip driven swimmers and multipole expansion as before.
For the slip driven swimmers the error decays again below $10^{-3}$ one radius away from the swimmers.
However, the multipole expansion gives much larger errors in the whole domain.
This is expected as the flow generated with the multipole expansion is purely additive, thus, the flow does not obey any boundary conditions on the squirmers surface.
In the slip driven swimmer case, however, the flow obeys the slip condition on their surface,
which renders a more accurate flow.
The error in the velocity field affects the trajectory of the squirmers.
We compare the trajectory of the well-resolved squirmers with those generated by the slip driven swimmer, with $N_b=162$ blobs, and by the multipole method
and show the difference versus time in Fig.\ \ref{fig:two_squirmers_trajectory}.
In both cases the error is low, at short times the error in the trajectory is linear in time and the slope increases for longer times.
However, the active slip swimmer has an error two orders of magnitude lower,
which one could have expected from the flow error shown in Fig.\ \ref{fig:two_squirmers}.

The computational cost of the slip driven swimmers can be made linear in the number of blobs using fast methods.
Thus, reducing the number of blobs by a factor $2562 / 162 \approx 16$  speeds up the low resolution simulations in the same proportion
without a significant error increase in the flow one radius away from the swimmers.
The multipole expansion method is the least computational expensive method, as it does not need to solve any linear system to apply boundary conditions.
However, as shown in the last panel of Fig.\ \ref{fig:two_squirmers}, that efficiency comes with a drop on the accuracy.
This example shows the advantages of the slip driven swimmer method to study suspensions of swimmers beyond the dilute regime,
or near obstacles, where hydrodynamic reflections play a role.

\subsection{Flow around a bacterium}
\label{sec:bacteria_flow}

\begin{figure}
  \begin{center}
    \includegraphics[width=0.4 \columnwidth]{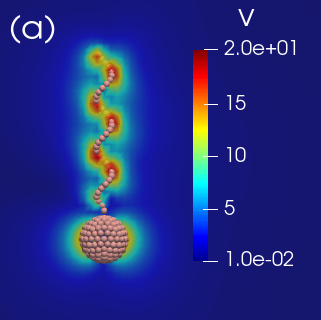}
    \includegraphics[width=0.4 \columnwidth]{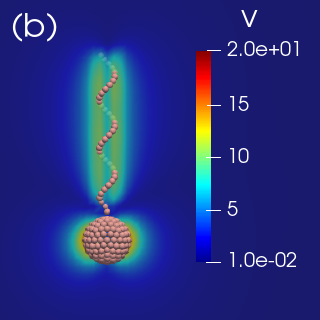}
    \includegraphics[width=0.4 \columnwidth]{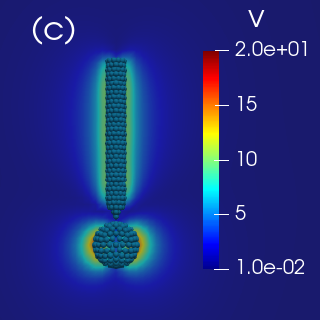}
    \includegraphics[width = 0.4 \columnwidth]{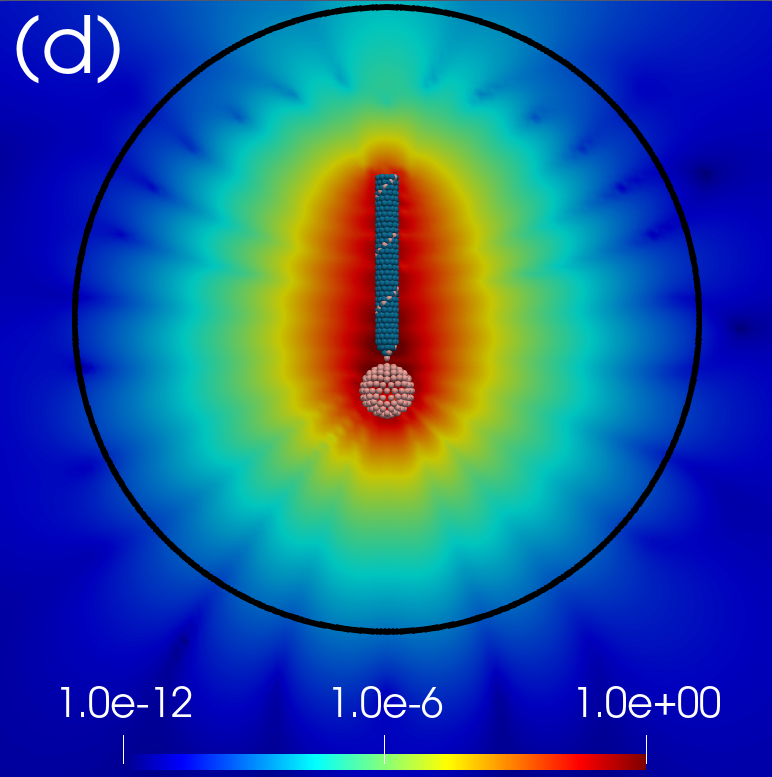}    
    \caption{Velocity field around microswimmers with swimming speed $U=2.55\, \si{\mu m / s}$.
      (a) Instantaneous flow field around a flagellated swimmer.
      Note that the flow velocity near the rotating bacterium is much larger than the swimming speed
      because the flagellum rotates fast.
      (b) Time averaged flow field of a flagellated swimmer in the swimmer frame of reference.
      (c) Time averaged flow field around a mapped swimmer.
      (d) The flow field difference between flagellated and mapped swimmers.
      The black circle represents the shell of radius $R_{\text{shell}}=12\,\si{\mu m}$ used in the optimization problem.} 
    \label{fig:bacteria_flow}
  \end{center}
\end{figure}

Here, we compare the flow field generated by a flagellated bacterium and a slip driven swimmer.
This test shows how our method allows to simulate the average flow generates by complex shaped swimmers.
We consider a bacterium with a spherical body of radius $R=1\, \si{\mu m}$ and a flagellum of length $L=10\, \si{\mu m}$,
see Fig.\ \ref{fig:bacteria_flow}a.
The flagellum is modeled as a rigid body with an helical shape, with wavenumber $k=2.86\,\si{\mu m^{-1}}$ \deleted{$k=2.8571$} and amplitude $\alpha = 0.35\,\si{\mu m}$,
attached to the body so it can only rotate along its axis \cite{Higdon1979}.
These geometric parameters are fixed throughout the paper unless otherwise mentioned.
We apply equal but opposite torques to the body and flagellum  so they rotate in opposite directions and the bacterium swim forward.
We set the magnitude of the torque so their relative angular velocity is $\omega = 63.46\, \si{rad / s}$ \deleted{$\omega = 63.4632\, \si{rad /s }$}
and the swimming speed is $u=2.55\, \si{\mu m/s}$\deleted{$U=2.5535\, \si{\mu m/s}$}.
To solve the hydrodynamic equations we use the rigid multiblob method for articulated bodies \cite{Usabiaga2022}.
The slip driven swimmer is modeled as a rigid body with the same spherical body as the bacterium and a tail that would envelope the flagellum,
see Fig.\ \ref{fig:bacteria_flow}c.
The active slip, still to be determined, acts on all the blobs belonging to the spherical body and tail.

The instantaneous flow field around the flagellated swimmer is shown in Fig.\ \ref{fig:bacteria_flow}a.
As explained before, matching the average flow allows to map the flagellated swimmer with a constant slip and therefore use large time steps
in simulations of slip driven swimmers.
For this reason, we compute the average flow during one flagellum rotation, see Fig.\ \ref{fig:bacteria_flow}b.
We also compute the average flow on a Chebyshev-Fourier grid of order $n=16$ (i.e.\ with $2(n+1)^2=578$ points)
defined on a shell of radius  $R_{\text{shell}}= 12\, \si{\mu m}$ surrounding the swimmer.
It is important that the average flows are computed in the body frame of reference and not the laboratory frame of reference as the swimmer will translate and rotate during one period.
The average velocity of the bacterium is also computed, which can be done from the instantaneous velocity as explained by Higdon \cite{Higdon1979}.

Using the average flow and the average velocity we calculate the optimum slip with the algorithm described in Sec.\ \ref{sec:map}.
The flow generated by the slip driven swimmer is shown in Fig.\ \ref{fig:bacteria_flow}c.
The difference between the two average flows is shown in Fig.\ \ref{fig:bacteria_flow}d.
It is evident from the figure that the near field error is of order $\mathcal{O}(1)$, however, the error decays fast with the distance.
At one body diameter away from the swimmer the error decays to about $10^{-3}$, one flagellum length away below $10^{-6}$ and
in the far field the error is of the order $\mathcal{O}(10^{-12})$.
The high accuracy of the flow at medium distances suggest that our scheme could be used to model bacterial suspensions beyond the dilute limit,
a regime where hydrodynamic interactions are not pairwise additive and therefore, where  multipole expansion methods are inaccurate.

\subsection{Bacterium above a wall}
\label{sec:bacteria_wall}

\begin{figure}
  \begin{center}
    \includegraphics[width=0.4 \columnwidth]{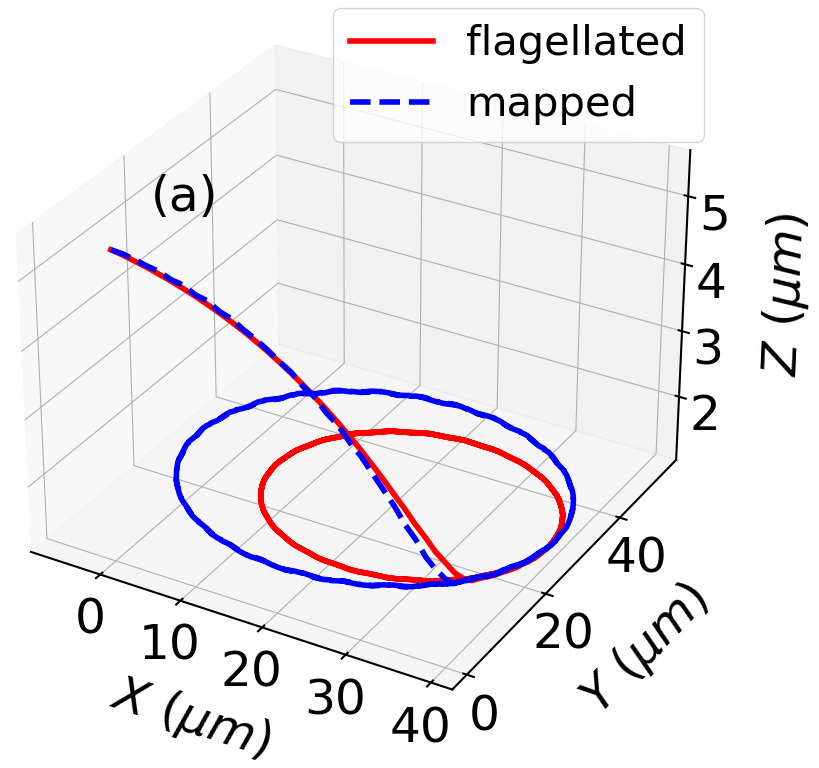}
    \includegraphics[width=0.4 \columnwidth]{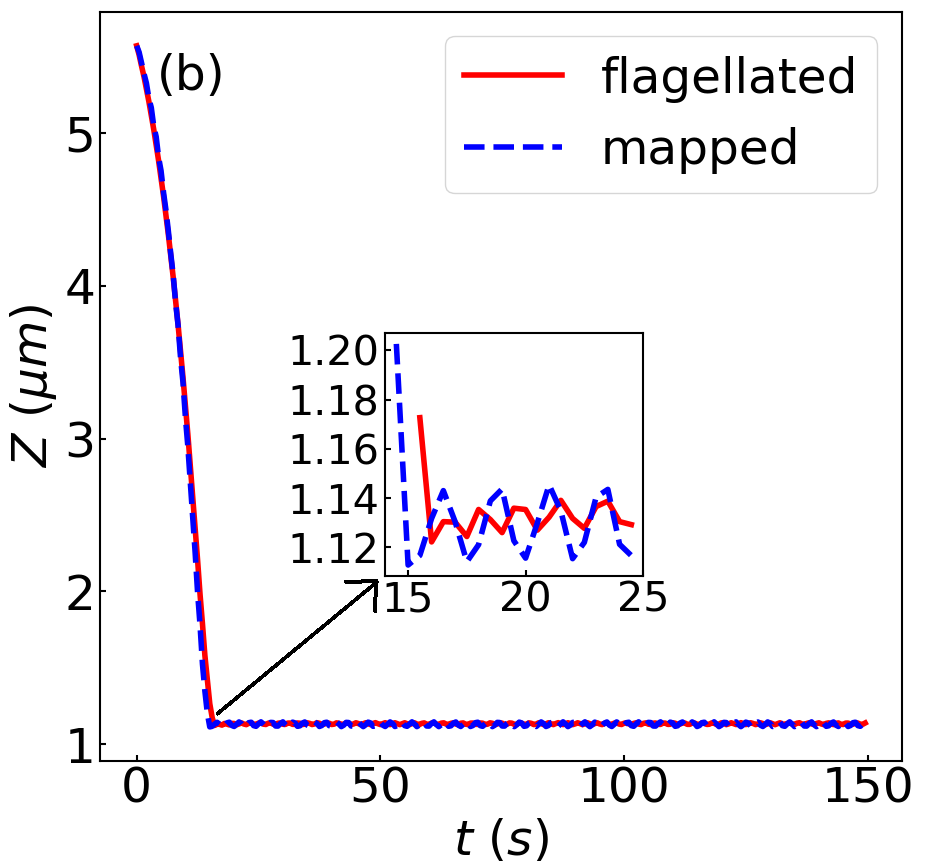}    
    \caption{(a) Circular trajectories, of radius $R_{\text{flag}}=17.0\, \si{\mu m}$ and $R_{\text{map}}=22.4\, \si{\mu m}$,
      of a flagellated bacterium and a mapped swimmer near a wall.
      (b) Temporal evolution of the height from the wall. The inset figure shows the zoomed temporal data when the swimmer reaches
      the wall and attains a steady height.
    }
    \label{fig:circular_traj}
  \end{center}
\end{figure}

In test we validate the active slip driven swimmers moving close to boundaries where hydrodynamic interactions are very important.
Our algorithm to determine the optimum slip, $\bu_s$, is designed to guarantee that the slip driven swimmer has the same average velocity as the flagellated swimmer.
However, as the computations to determine $\bu_s$ are done in bulk the swimmer velocity could be different near boundaries,
where other hydrodynamic interactions affect the dynamics.
One could compute an optimum slip near a boundary but the result would depend on the relative position and orientation between the swimmer and the boundary.
Thus, such approach will not be flexible to study suspensions of swimmers or swimmers near complex shaped obstacles.
Therefore, we prefer to use always the slip computed in bulk and study if our mapped swimmers can reproduce the dynamics of the flagellated ones near obstacles.
As microswimmers are very sensitive to confining environments, this is a hard test for our method.

It is well known that flagellated microswimmers follow circular trajectories in the counter-clockwise direction near a no-slip boundary \cite{Lauga2006}.
For this reason, we study here the dynamics of the bacterium from Sec. \ref{sec:bacteria_flow} near an infinite flat wall.
The bacterium is initialized swimming parallel to the wall at height $Z=5.6\, \si{\mu m}$.
In Fig.\ \ref{fig:circular_traj}a we show that both the flagellated and the slip driven swimmer approach the wall and then move along circular trajectories.
Furthermore, the steady state vertical distance from the wall in the case of mapped swimmer is in good agreement with the flagellated swimmer,
see Fig.\ \ref{fig:circular_traj}b.
However, the radius of the circular trajectory is $30\%$ larger for the slip driven swimmer. 
As the flow error near the swimmer was of order one it reasonable that the trajectory of our mapped swimmer is not exactly the same.
However, we consider the agreement reasonable good.
It is important to keep in mind that the trajectories are very sensitive to flagellar parameters such as its length,
helical radius or pitch \cite{Smith_PRS_2010,Kim_PoF_2022}.
For instance, a $50\%$ decrease in the flagellar length shows a $>80\%$ decrease in the radius of the circular trajectory \cite{Smith_PRS_2010}.
Thus, it is difficult to obtain a perfect agreement with any simplify model.

\subsection{Bacteria scattering}
\label{sec:bacteria_scattering}

\begin{figure}
  \begin{center}
    \includegraphics[width=0.65 \columnwidth]{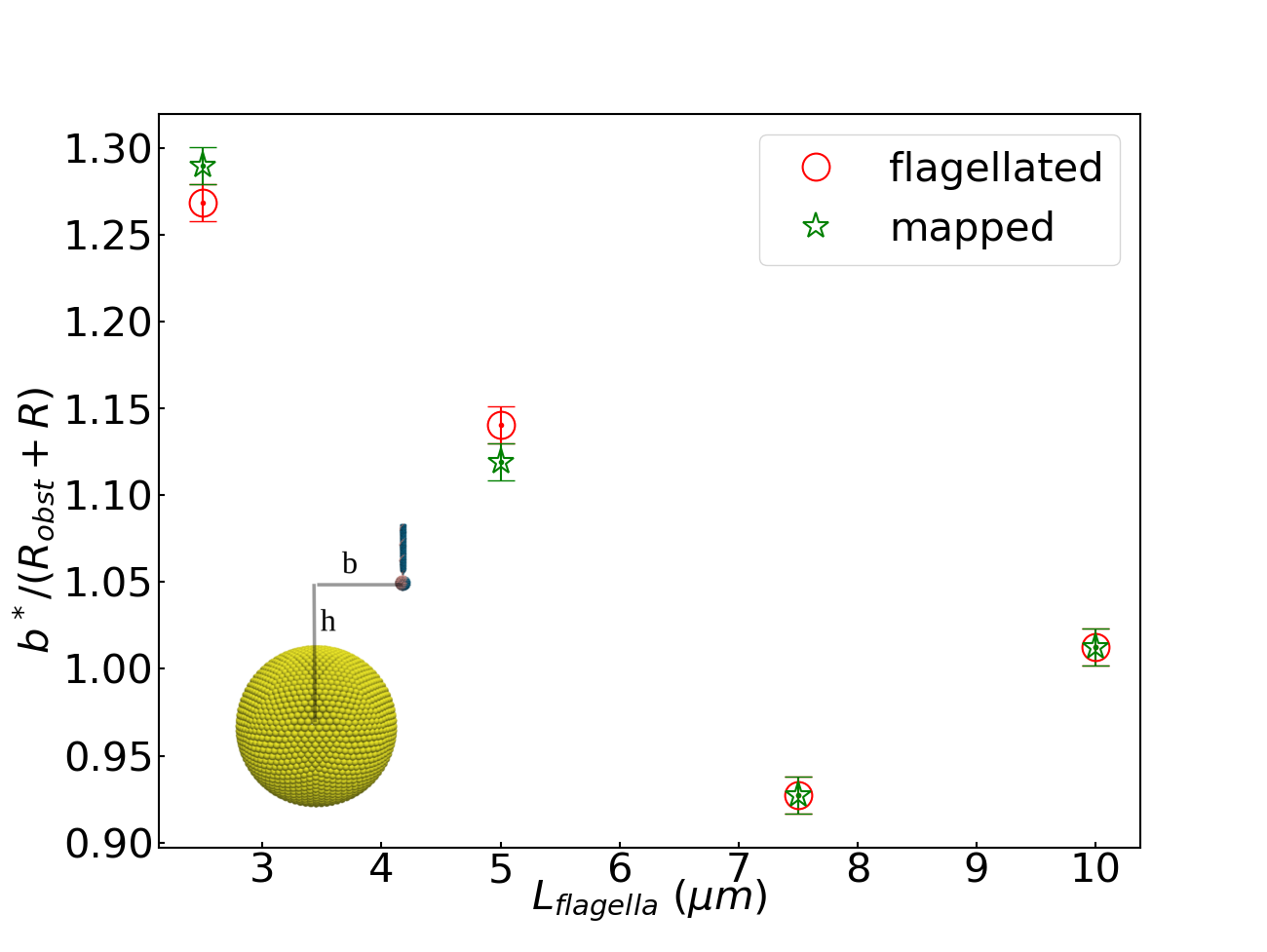}
    \caption{Critical impact distance with error bars, $b^*$, versus flagellum length, $L_{\text{flagella}}$.
      The results for flagellated swimmers, red circles, and mapped swimmers, green stars, show a good agreement.
    } 
    \label{fig:crit_dist_scattering}
  \end{center}
\end{figure}

We continue the study of interaction with boundaries by looking at the swimmer scattering with a fixed spherical obstacle of radius $R_{\text{obst}}=10.73\, \si{\mu m}$.
Without hydrodynamic interactions a swimmer moving towards an obstacle, with an impact parameter $b$, will collide whenever $b \le R_{\text{obst}} + R$,
where $R$ is the swimmer radius.
However, the hydrodynamic interactions can modify the critical impact parameter, $b^*$,
i.e.\ the distance below which the microswimmer hits the obstacle and above which it avoids the collision \cite{Slomka2020a}.
Here, we measure $b^*$ for flagellated and slip driven swimmers.
We use flagellated swimmers as the one from Sec.\ \ref{sec:bacteria_flow} but with several flagellum lengths, $L_{\text{flagella}}$.
For each bacterium we compute the optimum slip in bulk.
Then, we simulate the swimmers moving towards the obstacle from a distance $h = 90\, \si{\mu m}$ and with impact parameter $b$ to determine if they collide or avoid the impact.
We try several values of $b$ for each case to bound the value of the critical impact parameter $b^*$.
Figure \ref{fig:crit_dist_scattering} shows the variation of $b^{*}$ with $L_{flagella}$.
The comparison between flagellated microswimmer and mapped surface slip swimmer agrees very well for several flagellar lengths.
Thus, the simplified slip model can be used to study the interactions between obstacles and swimmers at a reduced computational cost.

\subsection{Two bacteria trajectory}
\label{sec:two_bacteria}

\begin{figure}
  \begin{center}
    \includegraphics[width=0.65 \columnwidth]{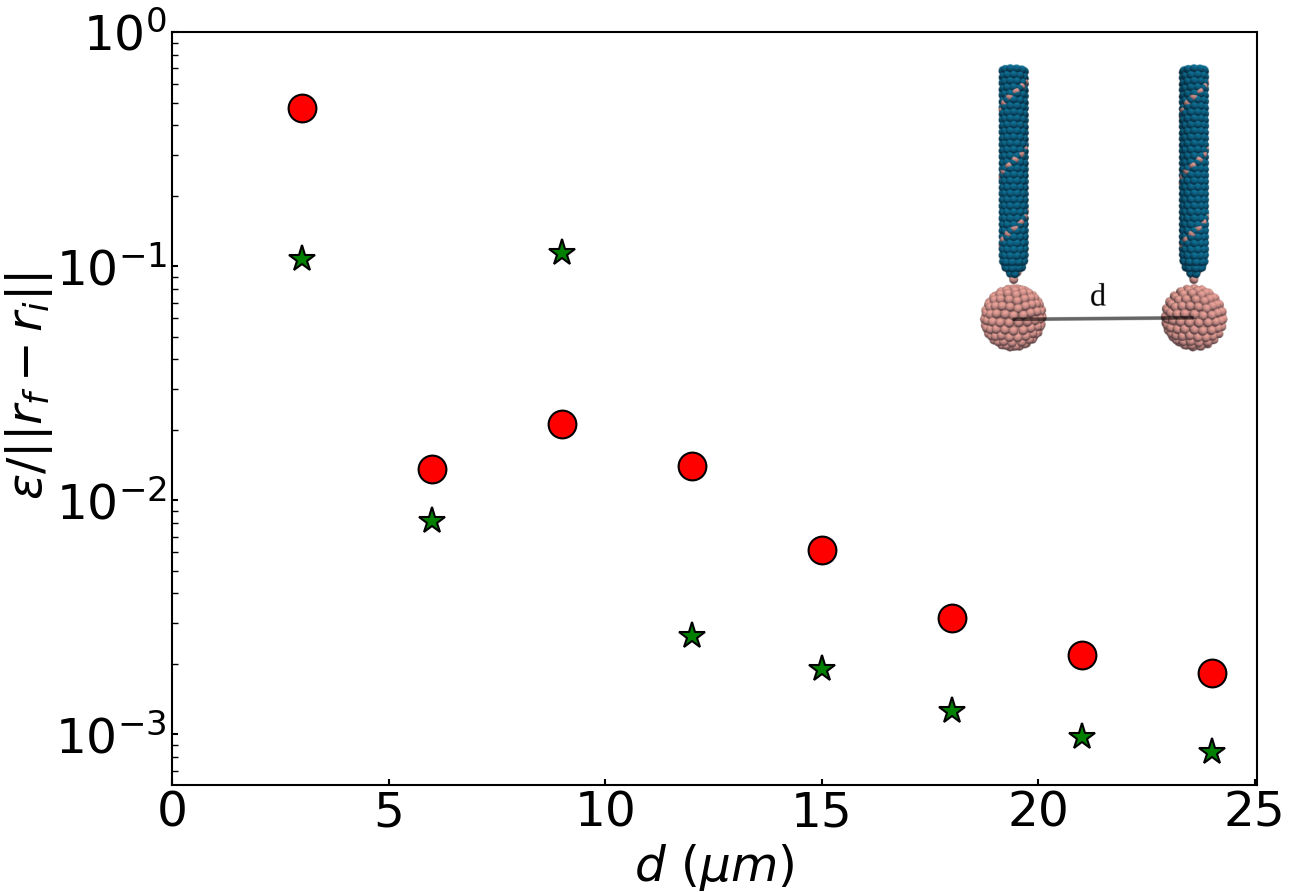}    
    \caption{Error in the position of one swimmer normalized with the its displacement versus the initial center to center separation distance, $d$,
      to a second swimmer.
      Results for flagellated versus mapped, red circles, and flagellated versus flagellated rotated $180^\circ$, green stars.
    } 
    \label{fig:error_vs_distance}
  \end{center}
\end{figure}

We carry out one more test to validate our scheme in situations where the hydrodynamic interactions with nearby objects are important.
We investigate here if the slip driven swimmers can recover the trajectory of two swimmers moving next to each other. 
In this test we set two flagellated bacteria, initially parallel and at distance $d$ (see inset in Fig. \ref{fig:error_vs_distance}), and let them swim for $300$ flagellum rotations
during which time they move about $30$ body radius.
Then, we repeat the simulation but using two slip driven swimmers.
We compare the final position and define an error as the difference between the position of one of the flagellated swimmers with one of the mapped swimmer.
We show the error, normalized with the swimmer displacement and for several initial distances $d$, as red circles in Fig.\ \ref{fig:error_vs_distance}.
When the initial distance between the swimmers is small, $d=3\,\si{\mu m}$, the error is of order one as expected.
Then, it decays for large distances to a relative error of about $10^{-3}$.
Although the error at large distances is small it is much larger than the error of the flow at large distances, $\mathcal{O}\pare{10^{-12}}$, or
the average swimming velocity error for a single swimmer, $\mathcal{O}\pare{10^{-15}}$.
To understand these large cumulative errors it is important to note that the trajectories are very sensitive to the details of the flagellum
and also that we are matching the average velocity but not the instantaneous velocity.
This second detail makes difficult to compare the trajectories pointwise in time
and can even affect the overall dynamics \cite{Walker2022}.

To verify the possibility of improvement, we run a second simulation with two flagellated swimmers
 where the flagella of one of them is initially turned half a rotation around its axis.
As consequence, the initial configuration is only slightly different than in the first case.
The difference between the final position of one bacterium from the first case and one from this simulation is shown as green stars in Fig.\ \ref{fig:error_vs_distance}.
One can see that the error is of the same magnitude when comparing with the slip driven swimmers.
We conclude that it is not feasible to significantly improve the accuracy, since the trajectories are so sensitive that a simple rotation
of the flagella leads to observable differences.

\section{Conclusions}
\label{sec:conclusions}

CFD is a well suited tool to study the swimming of microorganisms.
As different problems require a different balance between accuracy and computational cost there is not a single best numerical method to study these problems.
Here, we have presented a method that approximates the flow generated by flagellated swimmers by mapping the complex dynamics of their flagella to an active slip velocity.
This approach allows to use much larger time steps and in some cases less degrees of freedom to discretize a swimmer.
We have shown that mapping a swimmer in bulk is enough to recover its characteristic dynamics near obstacles.
There are many interesting applications for our scheme like the interaction of swimmers with obstacles and other swimmers, guided taxis and more.
We have shown a few examples here to showcase the flexibility and accuracy of the new method. 
Our scheme allows to model specific flagellated swimmers at a much reduced cost.

Our approach allows to introduce external forces acting on the swimmers but some considerations should be mentioned.
If the shape of the active slip swimmer matches the shape of the original swimmer,
its rigid motion under external forces and torques would generate the same flows than those created by the original swimmer under the same forces.
This could be the case of, for example, an ellipsoidal swimmer covered by cilia.
However, for complex shaped swimmers the shapes of the flagellated and the active slip swimmers may differ.
In that case, the flows generated by their rigid motions under external forces would also differ.
Therefore, the approach presented here may not be well suited to study, for example, the sedimentation of heavy complex shaped swimmers.
Nonetheless, this approach is still valid when the external forces are short ranged and only act briefly
or when the shapes of flagellated and active slip swimmers are similar enough.
For example, in the test of Sec.\ \ref{sec:bacteria_wall}, a bacterium swimming above a wall, we included steric interactions between the bacterium and the wall
and our approach was able to recover the original trajectory with reasonable accuracy.

The algorithm delineated in this paper can be adapted or extended in several ways.
The most immediate one is to employ a boundary integral method with spectral accuracy to simulate the slip driven swimmer.
This would allow to match the near field to a high accuracy at a reasonable computational cost \cite{Stein2022}.
Another possible extension using the rigid multiblob method is to optimize the blobs' location, and not only their slip, to minimize the flow error.
Such idea has been used to match the mobility of rigid bodies to a high accuracy and it could be used here \cite{Broms2022}.
Also, it is possible to include additional constraints in the linear system \eqref{eq:least_squares}.
For example, we could add a constraint to force the slip to be tangential to the swimmers surface.
We explored this idea but we did not find any relevant improvement in the accuracy of the flow.
Another interesting possibility is to model some parts of swimmer B as a porous material \cite{Abade2010a,Usabiaga2016}.
A porous media could be a better model, for example, for the envelop around a bacteria flagellum since the
rotating flagellum is not really a rigid body where the flow obeys the no-slip condition.

Our implementation of the inverse problem and the utilities to compute flows generated by swimmers are publicly available 
on GitHub 
(\url{https://github.com/stochasticHydroTools/RigidMultiblobsWall/tree/master/multi_bodies/examples/mapping}).

\vspace{1cm}
\noindent
{\bf Declaration of Interests:} The authors report no conflict of interest.

\section*{Acknowledgments}
The project that gave rise to these results received the support of a fellowship from ``la Caixa''
Foundation (ID 100010434), fellowshipLCF/BQ/PI20/11760014, and from the European Union’s Horizon 2020 research and innovation
programme under the Marie Skłodowska-Curie grant agreement No 847648. 
Funding provided by the Basque Government through the BERC 2022-2025 program
and by the Ministry of Science and Innovation: BCAM Severo Ochoa accreditation
CEX2021-001142-S/MICIN/AEI/10.13039/501100011033 and the project PID2020-117080RB-C55
``Microscopic foundations of soft matter experiments: computational nano-hydrodynamics (Compu-Nano-Hydro)'' are also acknowledged.


\bibliographystyle{elsarticle-num}
\bibliography{biblio}

\end{document}